\documentclass{aa}

\usepackage{graphicx}
\usepackage{txfonts}
\begin{document} 

   \title{A 52 hours VLT/FORS2 spectrum of a bright $z\sim 7$ HUDF galaxy:\\ no $Ly\alpha$ emission.
\thanks{This work is based on data collected at ESO VLT (prog.ID  084.A-0951(A),  086.A-0968(A), 088.A-1013(A)
and 088.A-1008(A)), and at NASA HST.}}


   \author{E. Vanzella \inst{\ref{Bol}},
           A. Fontana \inst{\ref{Rom}},
           L. Pentericci \inst{\ref{Rom}},
           M. Castellano \inst{\ref{Rom}},
           A. Grazian \inst{\ref{Rom}},
           M. Giavalisco \inst{\ref{UMASS}},
           M. Nonino \inst{\ref{Tri}},
           S. Cristiani \inst{\ref{Tri}}, \\
           G. Zamorani \inst{\ref{Bol}},
           \and
           C. Vignali \inst{\ref{unibo}}
          }
\institute{
INAF - Osservatorio Astronomico di Bologna, via Ranzani 1, I-40127 Bologna, Italy\label{Bol} \and
INAF - Osservatorio Astronomico di Roma, via Frascati 33, 00040, Monteporzio, Italy\label{Rom} \and
INAF - Osservatorio Astronomico di Trieste, via G. B. Tiepolo 11, I-34131, Trieste, Italy\label{Tri} \and
Department of Astronomy, University of Massachusetts, 710 North Pleasant Street, Amherst, MA 01003, USA\label{UMASS} \and
Dipartimento di Fisica e Astronomia, Universita' degli Studi di Bologna, Viale Berti-Pichat 6/2, 40127 Bologna, Italy\label{unibo}
}

   \date{Received -; accepted -}


  \abstract
  {}
   {We aim to determine the redshift of  {\it GDS\_1408}, the most 
    solid $z\sim7$ galaxy candidate lying in the Hubble Ultra 
    Deep Field.} 
   {We have used all the VLT spectra of {\it GDS\_1408} collected by 
    us and two other groups with FORS2 at VLT in the last five 
    years, for a total integration time of {\it 52hr}. The combined 
    spectrum is the deepest ever obtained of a galaxy in the 
    Reionization epoch.}
   {We do not detect any emission line or continuum over the whole wavelength 
    range, up to 10100\AA. Based on an accurate 
    set of simulations, we are able to put a stringent upper limit of 
    $f$($Ly\alpha$)~$<3 \times 10^{-18}~erg/s/cm^{2}$ at 3-9 sigma 
    in the explored wavelength range, corresponding to a rest-frame equivalent width 
    $EW<9$~\AA. Combining this limit with the SED modelling we refine the 
    redshift to be $z=6.82 \pm 0.1$ (1-sigma). The same SED fitting 
    indicates that {\it GDS\_1408} is relatively extinct ($A_{1600}\simeq 1$) 
    with a dust corrected star formation rate of $\simeq 20 M_{\odot} yr^{-1}$. 
    The comparison between the un-attenuated equivalent width predicted 
    by the case-B recombination theory and the observed upper limit,
    provides a limit on the effective $Ly\alpha$ escape fraction of 
    $f_{esc}^{eff}(Ly\alpha)<8$\%. Even though we cannot rule out a major 
    contribution of the inter/circum galactic medium in damping the line, 
    a plausible interpretation is that {\it G2\_1408} is moderately evolved and 
    contains sufficient gas and dust to attenuate the $Ly\alpha$ emission, 
    before it reaches the intergalactic medium.}
   {The redshift confirmation of even the best
    $z\simeq 7$ candidates is very hard to achieve (unless the 
    $Ly\alpha$ or unusually strong rest-UV nebular emission 
    lines are present) with the current generation of 8-10m 
    class telescopes. We show that both JWST and E-ELT 
    will be necessary to make decisive progresses. Currently, the increased redshift 
    accuracy obtained with this kind of analysis makes ALMA an interesting 
    option for the redshift confirmation.}


    \titlerunning{VLT ultradeep spectroscopy of a $z\sim7$ galaxy}
   \authorrunning{Vanzella et al.}

   \maketitle

%

\section{Introduction}

Understanding the process of reionization of the intergalactic medium in the early 
Universe and the nature of the first galaxies responsible for that process are among the 
most important goals of modern cosmology \citep[and references therein]{rob10}.
Thanks to the deep and  panchromatic data, such as GOODS, CANDELS and Hubble Ultra 
Deep Field (HUDF), great progress has been made in our ability to identify and subsequently 
confirm (spectroscopically) galaxies at $z< 7$.
The most prominent spectral feature in the UV rest-frame wavelengths probed by 
optical/NIR spectroscopy at $z\simeq 7$ is the $Ly\alpha$ emission line
(e.g., \citealt{vanz11}, V11; \citealt{pente11,pente14}, P11, P14; 
\citealt{sche12,schenker14}, \citealt{ono12}, \citealt{shibuya12}).
However, at $z> 7$ the situation is still challenging and at present only a handful objects 
are spectroscopically secured. There are at least two main reasons:
(1) Physical processes in the galaxies:
the $Ly\alpha$ emission is a resonant atomic transition very sensitive to dust attenuation,
and can be used for a diagnostic of the physical processes occurring within the galaxy 
\citep{giava96,atek13}, since its strength and velocity profile depend 
on the instantaneous star formation rate, gas and dust content, metallicity, kinematics, and 
geometry of the interstellar medium. Therefore an evolution of the average galaxy properties 
with time (gas and/or dust properties and/or ionizing emission) can make it intrinsically absent
at specific cosmic epochs.
(2) Reionization: the $Ly\alpha$ emission line may suffer of damping effect 
due to an increase of the neutral gas fraction in the intergalactic and/or circum galactic 
media \citep{miralda00, dijkstra11, treu12} or a increasing incidence of
optically thick absorption systems \citep{bolton13} as the ending phase of the 
reionization is approached. 
Another limiting effect was instrumental: efficient near infrared spectrographs 
($\lambda \gtrsim 1\mu m$) with multiplexing capabilities are needed to capture UV rest-frame 
features for many targets at once, and only recently they are becoming
available (e.g., VLT/KMOS, Keck/MOSFIRE, LBT/LUCI).

\begin{figure*}
\centering
\includegraphics[width=16.5 cm, angle=0]{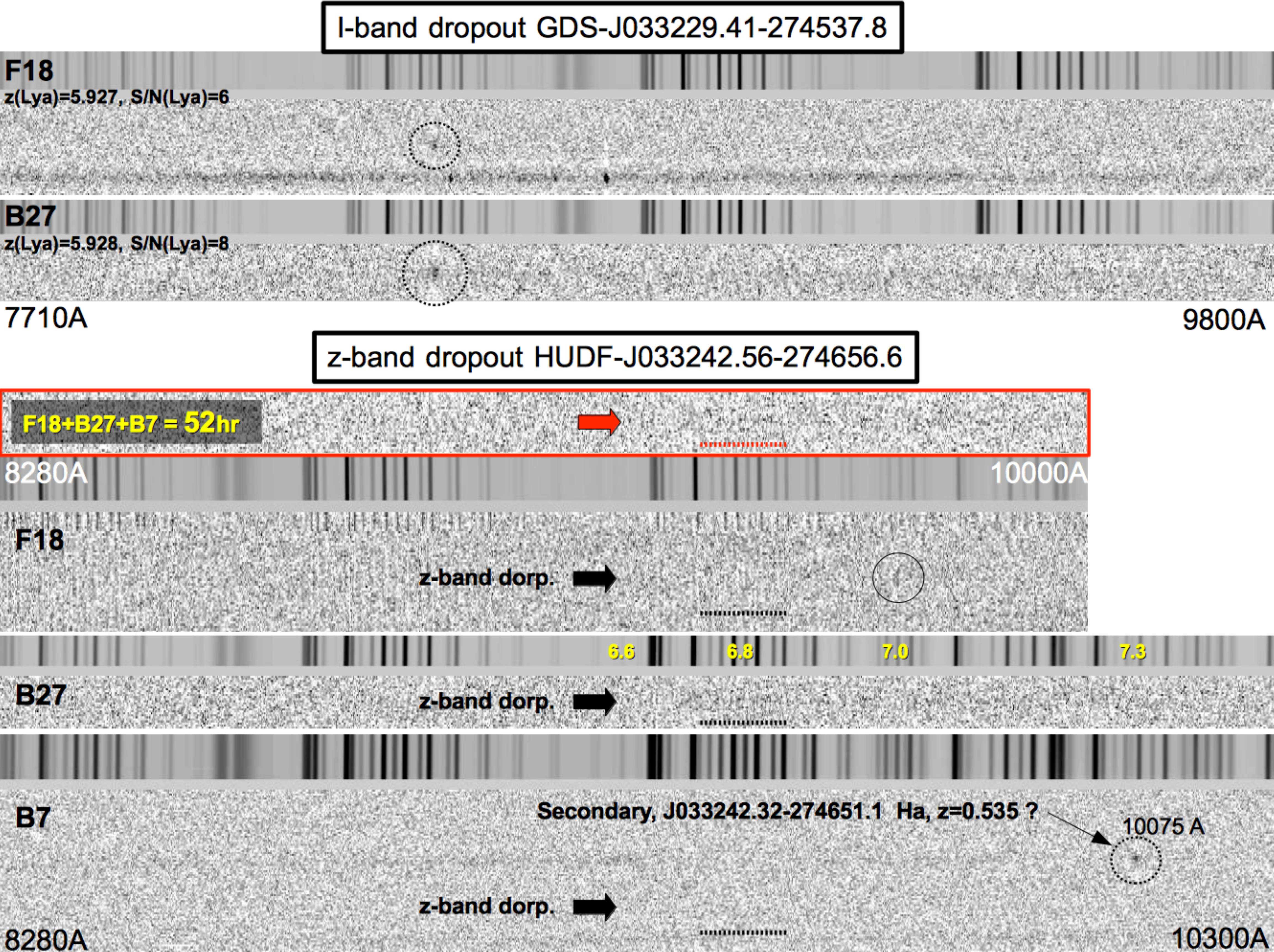}
\caption{Two dimensional signal to noise spectra and sky spectra of the galaxies discussed
in the text. {\bf Top:} spectra of the i-band dropout GDS~J033229.41-274537.8 observed in F18 (upper part) and
B27 (lower part). The $Ly\alpha$ line is marked with a dotted circle. {\bf Bottom:} the stacked {\it 52hr} spectrum
of the z-band dropout {\it G2\_1408} is shown, with the individual spectra of F18, B27, B7 (lower part). 
The expected position of the $Ly\alpha$ continuum break is marked with dotted horizontal lines (see text). 
In the B27 spectrum the redshift values are reported above the sky as a reference. In the F18 spectrum the 
dotted circle marks the older (here revisited) feature discussed in \citet{font10}. B7 spectrum shows also
the $H\alpha$ emission from the secondary object J033242.32-274651.1 at z=0.535, useful here
as an example of a still performing FORS+600z configuration beyond one micron.}
\label{spectra}
\end{figure*}
While point (1) is alleviated by the fact that (on average) 
the frequency of $Ly\alpha$ line emitters appear to increase as redshift increases 
\citep[$3<z<6.5$,][]{ouchi08,schaerer11,stark11}, the damping due to reionization may 
severely limit our current ability to confirm galaxies at $7<z<10$ \citep[but see][]{dijkstra10}.
\citet{fink13} find only one galaxy at $z=7.51$ out of 43 candidates at $z>6.5$ with the 
Keck/MOSFIRE and similarly \citet{schenker14} find only one possible $Ly\alpha$ line 
at $z=7.62$ in the their sample of 19 $z\sim 8$, remarking that some process is making 
the $Ly\alpha$ difficult to detect. 
Therefore, there is a large fraction of non-$Ly\alpha$ emitters
lying well within the first 
Gyr after the Big-Bang, whose nature is worth to investigate. 
Clearly, all these explainations assume that the efficiency of the color selection 
based on the $Ly\alpha$-break remains extremely high at $z\geq 7$, i.e. that most of current 
$z\geq 7$ candidates are indeed at their estimated redshifts.

In this work, we report on an ultradeep spectrum ({\it 52} hours integration)
of the brightest $z\simeq 7$ galaxy in the Hubble Ultra Deep
  Field ($F125W = 26.1\pm0.02$). Being by far the brightest $z\simeq 7$ candidate in this
  well-studied region, it has been continuously selected as a high
  redshift candidate from the earliest NICMOS to the current ultradeep
  HUDF data, at variance with other targets \citep{yan04, bou04,
    bou06, labbe06, bou08, oesch10, font10, mclure10, bunk10, yan10,
    fink10, cast10, wilkins11, bou11, grazian11, mclure13, bou14}. We
  focus here on the combination of ultradeep spectroscopy and
  photometry to derive new insights about its nature.

Errors are quoted at the $1\sigma$ confidence level, unless otherwise stated.
Magnitudes are given in the AB system (AB~$\equiv 31.4 - 2.5\log\langle f_\nu / \mathrm{nJy} \rangle$).
We assume a cosmology with $\Omega_{\rm tot}, \Omega_M, \Omega_\Lambda = 1.0, 0.3,
0.7$ and $H_0 = 70$~km~s$^{-1}$~Mpc$^{-1}$.

\section{FORS2 Observations and Data Reduction} \label{data}
HUDF-J033242.56-274656.6 galaxy ({\it G2\_1408} in \citealt{cast10} and hereafter) 
has been observed through four different VLT/FORS programs collected in the period 2009-2012, 
084.A-0951(A) (P.I. A. Fontana),  086.A-0968(A) -- 088.A-1013(A) 
(P.I. A. Bunker), 088.A-1008(A) (P.I., R. Bouwens), and 283.A-5063 (P.I., M. Carollo)
with exposure times of 18 (Fontana, F18), 27 (Bunker, B27), 7 (Bouwens, B7) and 8 (Carollo, C8) 
hours on target, respectively, for a total of $\simeq$60 hours integration time.
Unfortunately, in the program 283.A-5063 the {\it G2\_1408} source was placed at a position 
where the CCD has a defect and no
dithering has been performed, therefore we exclude the 283.A-5063 program from the 
following analysis. The total usable exposure time is 52 hours.
The median seeing was $\sim$0.8 arcsecond in all runs.
The F18 data were presented in \citet{font10} where
we reported a tentative detection of a $Ly\alpha$ line at z=6.972.
The other runs, B27 and B7, have been obtained subsequently
in many different nights (especially B27), and particular care has been
devoted to the alignments of the frames by using
bright sources and sky emission lines. 
The B27 program has also been presented in \citet{caruana14}, where
they reported a $S/N\simeq 3.2$ at the location of the putative $Ly\alpha$ at $z=6.972$.
We discuss in very detail any possible presence of $Ly\alpha$ emission 
by combining all the available programs.

Data reduction has been performed as in \citet{vanz11}, with particular
care to the sky subtraction. The classical ``A-B'' dithering scheme that combines 
the partial frames (A-B) and (B-A) is performed, with an additional treatment
that {\it equalizes for} local differences in the number counts between frames
(e.g., due to time variation of sky lines, distortion, etc.). 
The algorithm implements an ``A-B'' sky subtraction
joined with a zero (e.g., median) or first order fit of the sky
along columns that regularized possible local differences in the sky counts among 
the partial frames before they are combined.
Finally, the two-dimensional spectra have been combined with a weighted average,
and the subsequent resulting spectrum has been flux and wavelength calibrated. 
The two dimensional sky-subtracted
partial frames are also combined (in the pixel domain) to produce the 
weighted RMS map, associated to the final reduced spectrum. 
This allows us to calculate the two dimensional signal to noise spectra, 
useful to access the reliability of the spectral features 
(as we address below with simulations).

Further checks have been performed on other targets placed in the masks and on those 
in common between F18 and B27.
In particular, \citet{caruana14} targeted the same faint $i-$band
dropout we confirmed 
previously with F18 \citep{font10}, GDS~J033229.41-274537.8 at $z=5.927$.
While in the F18 spectrum the continuum is not detected, in our B27
spectrum we find a very faint trace redward of the line and a $Ly\alpha$ with estimated flux
of $3.8 \times 10^{-18}~erg/s/cm^{2}$ at S/N$\simeq 8$, about a factor 1.25 higher than F18,
whose flux is estimated to be $\simeq 3.0 \times 10^{-18}~erg/s/cm^{2}$ with S/N$\simeq 6.5$.
Both flux and error estimates are compatible within the flux calibration accuracy and 
different integration time.
The top part of Figure~\ref{spectra} shows the S/N spectra of the $z=5.927$ galaxy
(i.e., the reduced spectrum divided by its RMS map).

Following the above approach, we have combined the 52 hours of observations of {\it GDS\_1408}.
The three individual S/N spectra and the conbined one are shown in the bottom part 
of Figure~\ref{spectra}. The 52 hours spectrum is the deepest spectroscopic observation of 
a $z\sim 7$ galaxy obtained to date.

\section{Results}\label{results}

\begin{figure}
\centering
\includegraphics[width=8.5 cm, angle=0]{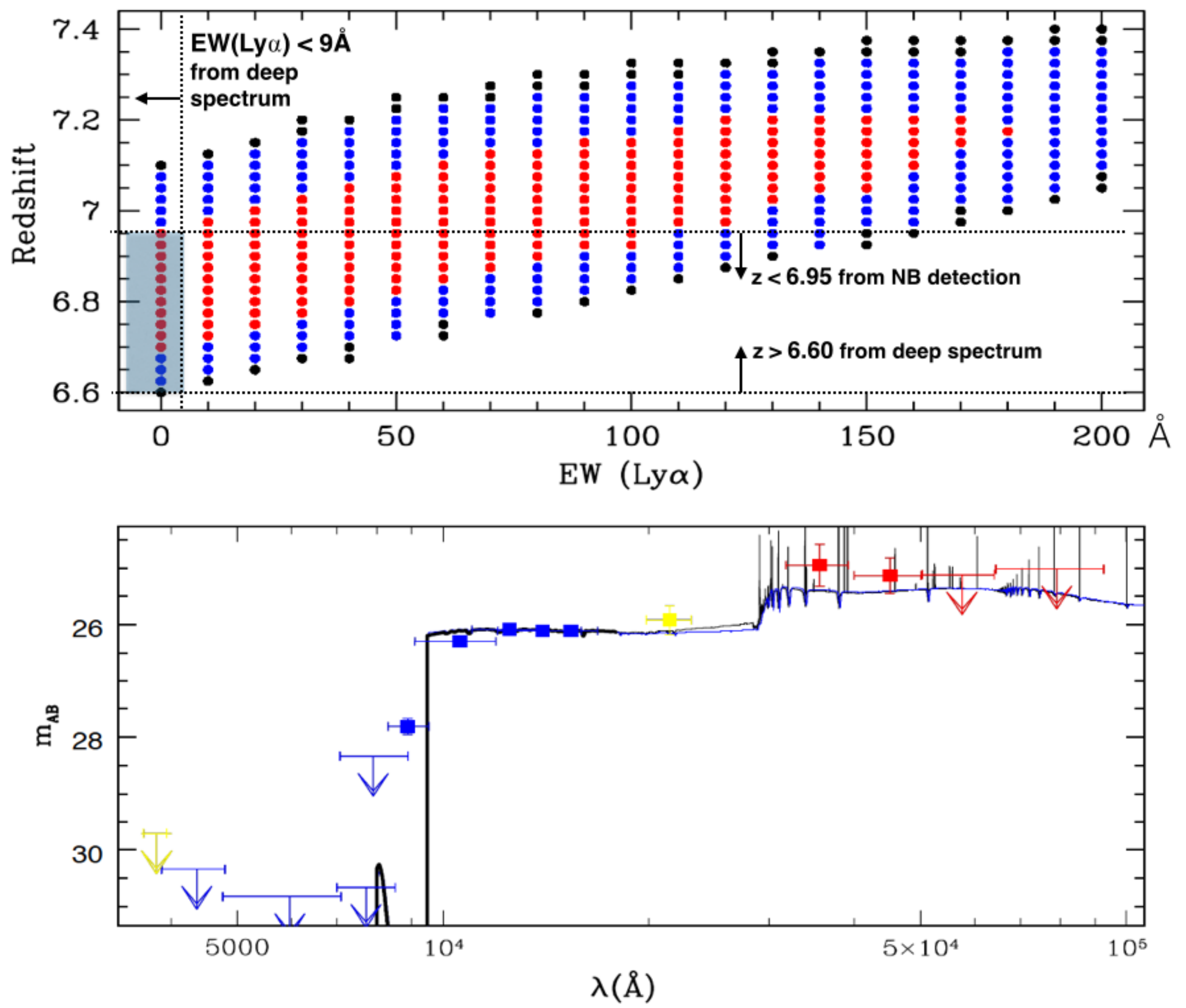}
\caption{{\bf Top:} The projected Redshift-$EW$($Ly\alpha$) solutions of the SED fitting
are shown. Black, blue and red symbols represent 3-,2-, and 1-sigma solutions. The well 
detected signal in the $z_{850}$-band is reproducible with large $EW$($Ly\alpha$) at higher 
redshift ($z>7$) or very faint $Ly\alpha$ emission at the lower redshift, $z<7$. 
Upper and lower horizontal dotted lines mark 
the limits provided by the NB973 detection and non-continuum detection below $z=6.6$ in the deep
spectrum, respectively. The vertical dotted line marks the upper limit of the $Ly\alpha$ 
rest-frame equivalent width. 
The transparent gray region underlines the most probable redshift interval.
{\bf Bottom:} SED fitting performed with BC03 
libraries with (thin black) and without (blue) nebular emission lines are shown.
The fit to the HST bands only is shown with the thick black line.
Blue, yellow and red points represent photometry from HST, VLT (U and K bands)
and Sptizer/IRAC, respectively. Upper limits at one sigma are marked with arrows.
}
\label{degene}
\end{figure}

\begin{figure}
\centering
\includegraphics[width=8.5 cm, angle=0]{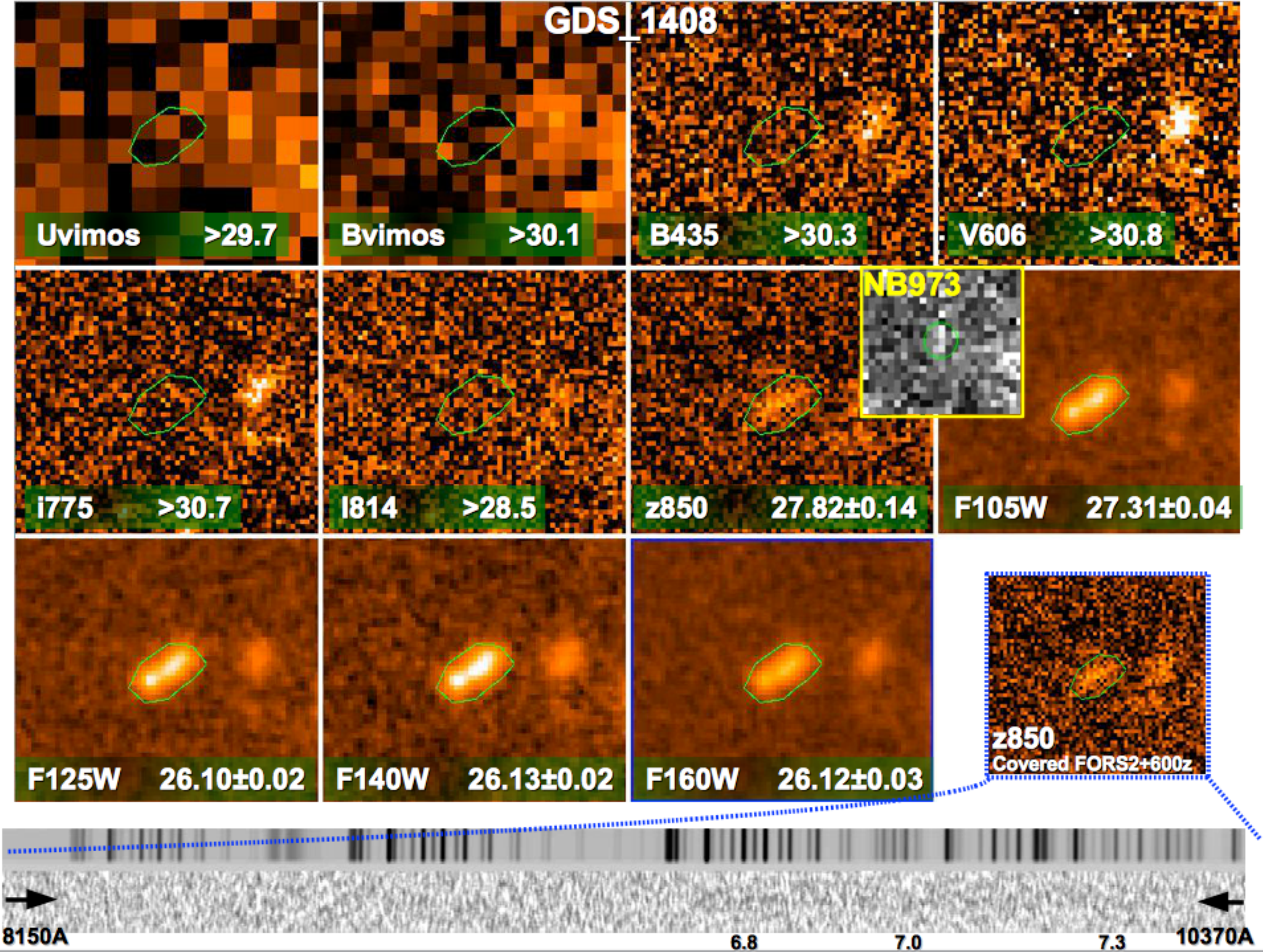} 
\caption{The cutouts ($3 ''\times 3''$) from U to H-band of {\it G2\_1408} are shown with the magnitudes
(limits are at one-sigma). The green contour marks the galaxy's shape as a guidance.
The inset image placed between the $z_{850}$ and F105W refer to the narrow band filter NB973 in
which the source is detected (3-sigma). The bottom-right cutout shows the clear
detection in the $z_{850}$ band ($S/N\sim 8$), for which there is a good coverage of the
ultradeep FORS2 spectrum (also reported in the bottom).
}
\label{HST}
\end{figure}

Before discussing any feature in the spectrum we note that the
available photometry already constrains the redshift in the range
($6.5<z<7.0$) where FORS2 is an efficient instrument for the
detection of the corresponding Ly$\alpha$, both in terms of wavelength
coverage and response.  First, we recall that globally the high
redshift nature ($z>6.5$) is guaranteed by the large observed break
between the ultradeep optical and near infrared bands ($\Delta m \simeq 4$
magnitudes) and the well determined flat behaviour of the SED in the
near infrared bands, detected with a high SNR (HST-WFC3, S/N$\simeq$20-50), 
as shown in Figure~\ref{degene} and ~\ref{HST}. This makes {\it GDS\_1408} one of
the most robust $z\sim7$ candidates.  Second, a reliable upper limit
on the redshift is provided by the clear detections in the $z_{850}$
band ($z<7.3$) and narrow band filter, NB973 ($z<7$), 
centered at $\lambda = 9755$\AA~and
$d\lambda=200$\AA~(Figure~\ref{HST}).
The limit provided with the $3\sigma$ detection in the NB973 is
 $z<6.94$ if the entire NB973 filter is capturing the galaxy continuum, 
$z_{MAX}=(9755-100)/1215.7-1=6.94$ (see inset of Figure~\ref{HST}).

\subsection{A deep upper limit to the $Ly\alpha$ flux}

Given the above upper limit $z_{MAX}$, the FORS2+600z configuration provides a safe 
constraint on the 
$Ly\alpha$ line flux.\footnote{As an example Figure~\ref{spectra} 
shows the secondary object in B7 spectrum 
in which the $H\alpha$ line is detected at z=0.535 consistently with the $zphot=0.5$, 
and showing that the FORS2+600z configuration is still performing well at $\sim$10100\AA.}
Indeed, there are no obvious spectral features in the three S/N spectra at the
position of the {\it G2\_1408} (marked by arrows), neither in the combined one 
(see Figure~\ref{spectra}).

We note that these deeper data do not confirm the tentative 
detection of a weak line located at $\lambda = 9691.5 \pm 0.5$\AA, corresponding to a redshift 
of 6.972 that we reported in \citet{font10} with $S/N<7$ (and shown with a
circle in the F18 spectrum of Figure~\ref{spectra}).
Exploiting the RMS map we derived, the reliability of the spectral feature
in the F18 spectrum  turns out to have $S/N \simeq 4.5$.
In the combined ({\it 52h}) spectrum, however, the S/N at the same wavelength position 
is even smaller, $\lesssim 3$, and suggests that the earlier tentative detection 
was most probably a noise spike.

To assign a statistical significance to our {\it 52h} non-detection, 
we estimated the minimum line flux 
reachable with the deep FORS2+600z spectrum by computing simulations as in 
V11 and P14. Two dimensional asymmetric $Ly\alpha$ lines have been
inserted in the science raw frames moving the line from $z=5.7$ to 7.3 
with $dz=0.0013$ (i.e., one pixel at the given spectral resolution), including 
the dithering pattern, varying the FWHM and the line flux, convolving with 
spectral resolution along the dispersion and with the seeing along the spatial
direction (extracted from the header of each science frame).
Knowing the exact position of the inserted lines in the raw frames 
(that by definition include also the cosmic rays), 
and including the full reduction pipeline process and the 
response curve, we can reliably access the limits attainable by the instrument.
The resulting S/N of the simulated lines are fully compatible with those
we observed at redshift 6 and 7 (P11, P14 and V11).

\begin{figure*}
\centering
\includegraphics[width=16.5 cm, angle=0]{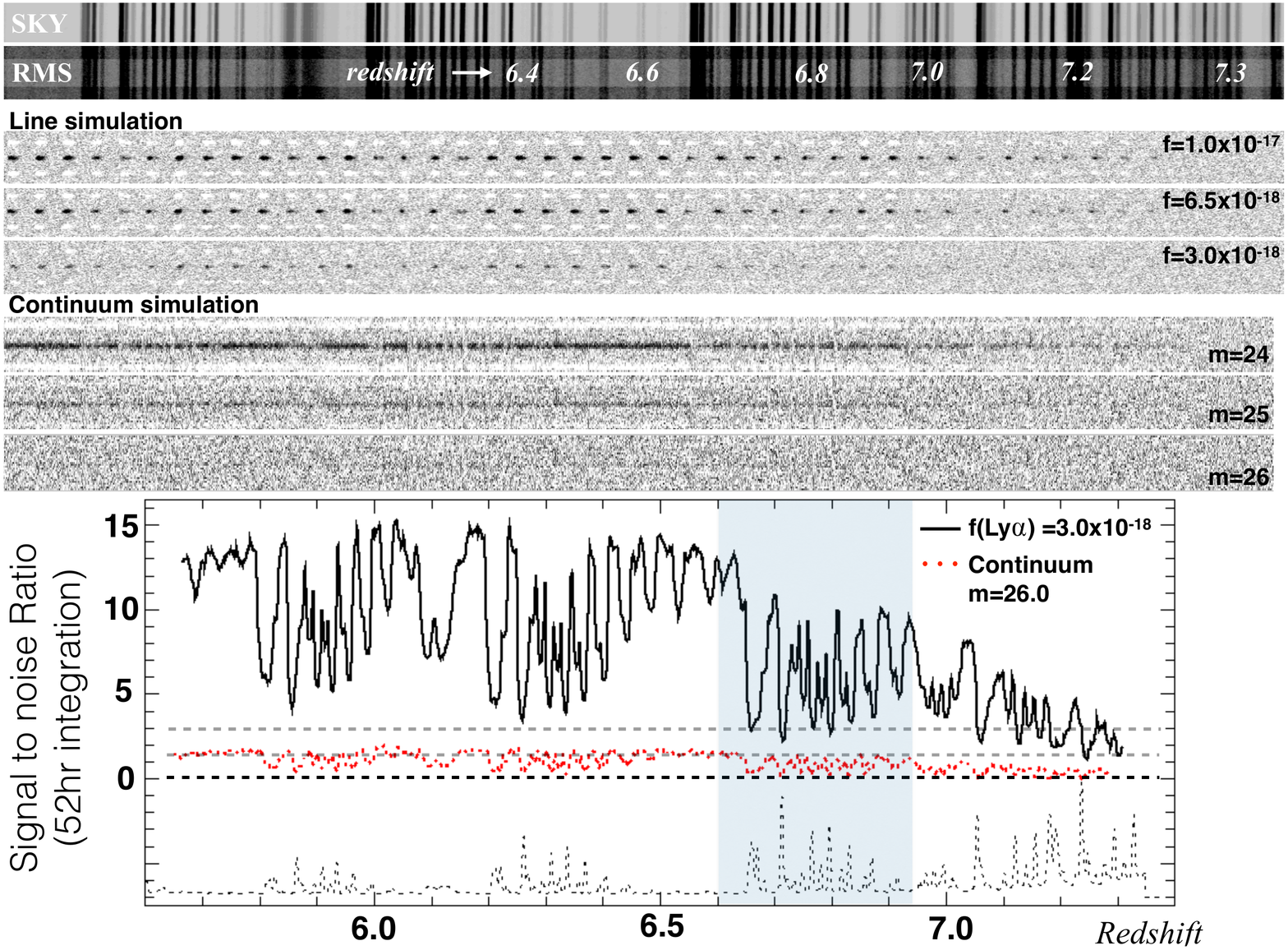} 
\caption{Simulation of the expected signal to noise ratio for emission
  $Ly\alpha$ line and continumm with 52 hours integration with FORS2. 
  From top to bottom: the two dimensional
  spectra of the sky and RMS map are shown as an example and
  guidance for the eye. Simulated signal to noise 2d-spectra of 
  line emission (snapshot) and continuum emissions are shown, with 
  indicated the line flux and magnitude at the (flat) continuum on 
  the right. It is clear that the S/N is lower where the sky lines are stronger. 
  In the bottom part the S/N ratio of the $Ly\alpha$ feature
  versus redshift (solid  black line) and the continuum 
  (red dotted line) are reported for the case of {\it GDS\_1408}.
  The transparent grey region marks the $z_{MIN}<z<z_{MAX}$ interval, see text for
  more details. Grey horizontal dotted lines mark $S/N=1.5$ and 3.0.}
\label{SIM_CONT}
\end{figure*}

The upper limit we derive from the combined {\it 52h} spectrum is 
$f$($Ly\alpha$)~$<3 \times 10^{-18}~erg/s/cm^{2}$ at 3-sigma 
(in the sky-lines) and up to 9-sigma (in sky-free regions) in the whole 
wavelength range, i.e., up to $z=7.0$ (see Figure~\ref{SIM_CONT}). 
Adopting the F125W=26.10$\pm$0.02 as the estimation of the continuum
under the $Ly\alpha$ line,
and given the relatively flat UV slope in the near infrared bands, the limit on the
$Ly\alpha$ flux corresponds to an upper limit on the equivalent width $EW<9$~\AA, 
with the same statistical accuracy.
It represents the faintest limit on the $Ly\alpha$ flux ever 
derived at $z>6.5$.

While absorption lines are clearly impossible to detect in our spectra, 
it is worth to investigate whether the non detection of the continuum 
in the {\it 52h} spectrum 
is compatible with the expected magnitude $\simeq 26.1$.
 To this aim, a set of
two-dimensional continua (with flat UV slope) without any absorption line
have been added to the raw frames as performed for line simulations, 
adopting a Gaussian profile in the spatial direction (consistent with
the observed seeing). They have been
added to the raw frames with dithering and dimmed in magnitude, 
from 23.0 to 27.0 with $dm=0.25$ (in all
the process the response curve has been taken into account). 
Figure~\ref{SIM_CONT} shows the results. 
The S/N decreases accordingly with the magnitude dimming and in the 
presence of sky emission lines, becoming impossible to detect at 
magnitude $\simeq 26.0$ unless the object is at redshift below 6.6
(see Figure~\ref{SIM_CONT}).
Both simulations on $Ly\alpha$ line and continuum show the clear decrease
of the S/N at the position of sky emission lines, i.e., it is lower where the
sky emission is stronger.
The results of the continuum simulation are in line with the observed faint galaxies, 
in particular with those we confirmed in P11 at $z\sim 6$ based on 
continuum-break only.

\subsection{Refined redshift}
Depending on its EW, the $Ly\alpha$ emission can contribute significantly
to the broad band photometry and as a consequence to the photometric
redshift determination.
In particular, the flux observed in the $z_{850}$ band (see Figure~\ref{HST}) depends on the
position of the $Ly\alpha$-line/break (i.e., the redshift), the IGM attenuation, the 
EW of the possible $Ly\alpha$ emission ($EW(Ly\alpha)$), and mildly on the UV slope of the
source, $\beta$ ($F_{\lambda}=\lambda^{\beta}$).
With the depth and photometric quality available for this galaxy and
the upper limit derived above, the redshift value can be refined.  To
this aim we performed the SED fitting with \citet{BC03}(BC03,
hereafter) templates by including different equivalent widths of the
$Ly\alpha$ (0-200\AA~rest-frame) and focusing solely on the break
between the optical and near infrared bands, i.e., $B_{435}$, $V_{606}$, $i_{775}$, 
$z_{850}$, $F105W$, $F125W$, $F140W$ and $F160W$.  We have not included the NB filter
  in this exercise since its photometry is not as accurate as the HST
  data. The templates adopted a range of $e$--folding star--formation
  timescales ranging from $0.1$ to 15 Gyr (the latter being in
  practice a constant-star formation history), metallicities from
  $Z=0.2Z_{\odot}$ to $Z=Z_{\odot}$, and a Calzetti attenuation curve
  with $A_V=0-3$. Figure~\ref{degene} shows the {\it plane}
  (redshift--$EW(Ly\alpha)$), in which the degeneracy is clear, i.e.,
  a progressively stronger $Ly\alpha$ emission is required at increasing
  redshifts to compensate for the increasing absorption by the IGM.
The degeneracy with the $Ly\alpha$ line is broken by including the
result from the ultradeep {\it 52h} spectrum discussed in Sect.~3.1, that
forces the $EW(Ly\alpha)$ to be smaller than 9\AA~rest-frame.
Once the Ly$\alpha$ constraints are included, the redshift of
  {\it G2\_1408} is constrained within $6.7$ and $6.95$ at $1\sigma$
  and $6.6$ and $7.1$ at $2\sigma$.
The top panel of Figure~\ref{degene} shows the resulting region
where the redshift would lie, that is $z=6.82_{-0.1}^{+0.1}$ (at one sigma).

We note that the $2\sigma$ range is consistent with the continuum-only 
simulations described above and the  non detection in the sky-free region of the 
spectrum at $9000$\AA$<\lambda<9280$\AA, {\it that both} suggest a lower limit 
of the redshift, $z_{MIN}=6.6$. Less significant, but consistent with $z_{MIN}$, is the 
slightly brighter 3.6$\mu m$ magnitude to respect the 4.5$\mu m$ channel in the data 
we have, that could 
suggest a flux boost of the [OIII]4959-5007+$H\beta$ structure in the first channel, while 
the $H\alpha$ line is outside (redward) the second one, respectively.
Indeed, this is confirmed with deeper IRAC data, in which
a clear break has been measured, [$3.6 -4.5$]$\mu m = 0.66 \pm 0.2$
(Smit et al.  in preparation, \citet{labbe13}). This is fully consistent with 
our redshift estimation.

The SED fitting with and without the nebular component has then been performed 
including the IRAC and ground based photometry.
We have included the full treatment of the nebular emission (both in lines and continuum), 
computed using the \citet{SB09} model, as described in \citet{cast14}.
The resulting SED fits with HST-only bands and the whole photometry are shown in the bottom 
panel of Figure~\ref{degene}. 
The derived physical quantities in the various cases agree within a factor of two.
This exercise provides an estimate  
of the stellar mass of $M=5_{-2}^{+3}\times10^{9}M_{\odot}$, an age $\simeq 0.1_{-0.05}^{+0.15}$ Gyr, 
a dust attenuation E(B-V)~$\simeq 0.1 \pm 0.05$ and a dust corrected star 
formation rate of $21_{-10}^{+20} M_{\odot}yr^{-1}$.

\section{Discussion}\label{dis}

The combination of excellent photometry and ultradeep
spectroscopic data lead us to conclude that the lack or
extreme weakness of a $Ly\alpha$ emission is a real feature of this
object. It is therefore worth investigating the reasons why the
$Ly\alpha$ line is not present.

\subsection{The effective $Ly\alpha$ escape fraction}
\citet{dijkstra13} differentiate between the term 'escape' fraction 
and 'effective escape' fraction of $Ly\alpha$ photons, $f_{esc}(Ly\alpha)$ and $f_{esc}^{eff}(Ly\alpha)$,
respectively. The former being related to the
transport of photons out of the galaxy's interstellar medium, while the latter 
includes further damping by the IGM, giving rise to a low-surface-brightness $Ly\alpha$ 
glow around galaxies. Therefore, $f_{esc}^{eff}(Ly\alpha)$ can be much 
smaller than $f_{esc}(Ly\alpha)$.
If we assume ''normal'' metallicities and IMFs, and that on average star 
formation is ongoing at equilibrium ($age>100Myrs$ with constant SFR), an estimation 
of the $Ly\alpha$ luminosity can be obtained imposing SFR(UV)/SFR($Ly\alpha$)$\simeq$1
\citep{hayes11,ver08}. Assuming the SFR(UV) corrected by
dust extinction is representative of the total ongoing star formation activity,
the intrinsic $Ly\alpha$~flux turns out to be $\simeq 4 \times 10^{-17} erg s^{-1}cm^{-2}$,
i.e., $EW(Ly\alpha)$ rest-frame $\simeq 120$\AA.
The upper limit on the observed $Ly\alpha$ line flux derived above corresponds 
to an observed $EW(Ly\alpha)<9$\AA~rest-frame (at $3-9\sigma$), 13 times smaller
than the expected intrinsic emission, that gives an upper limit 
of $f_{esc}^{eff}(Ly\alpha)<8$\%. 

\subsection{What is attenuating the $Ly\alpha$~emission ?}

The escape of $Ly\alpha$ photons from a gaseous and dusty interstellar
medium is a complex process which depends sensitively on a number of physical 
properties, such as dust and gas mass, star formation rate, metallicity, kinematic, 
escaping ionizing radiation, as well as the gas geometry and filling factor and the 
galaxy orientation \citep{neu91, ver08, ver12, hayes11, yajima14, laursen13, dijkstra14}.
Additionally, the neutral hydrogen in the IGM 
can scatter part of the $Ly\alpha$ photons and decrease the line luminosity. 
For example, \citet{laursen11} suggested that the 
average IGM transmission could be $\simeq$20\% at $z = 6.5$. 
Also in this case the transmission depends sensitively on the viewing 
angle and the environments of the galaxy, as it is affected by the 
inhomogeneous filamentary structure of IGM. At $z>6.5$ the IGM
damping could be the dominant effect. 

It is not possible to investigate in detail the above quantities with
the current information. Therefore we have to rely on global properties, e.g., 
by performing a comparative analysis with lower redshift galaxies at $z>3-4$.
First, it has been shown that
$2<z<6$ UV-bright star-forming galaxies ($L>L^{\star}$) show on average a deficit 
of $Ly\alpha$ emission with respect the fainter UV counterparts 
(e.g., \citealt{ando06}; \citealt{vanz09}; \citealt{lee13};
\citealt{balestra10}), explained by the fact that on average galaxies that 
are intrinsically brighter in the UV are also more star forming and more massive, 
more chemically enriched and thus more likely to be obscured by dust,
especially the $Ly\alpha$ resonant transition that is expected 
to be efficiently absorbed.  A decreasing
trend of $f_{esc}(Ly\alpha)$ emission with increasing E(B-V) has also
been observed in several studies \citep{giava96, atek08, ver08, kornei10, hayes11}.
The observed UV slope ($\beta = -2$) of {\it GDS\_1408}
suggests there is a dust attenuation $A1600\sim 1.3$
\citep{cast14, debarro14}. The inferred $SFR \simeq 21 M_{\odot}yr^{-1}$
and assuming the Schmidt law \citep{schmidt59,  kenn98a},
implies a non negligible amount of 
gas, $M_{gas}\simeq 3 \times 10^{9} M_{\odot}$, comparable to the
stellar mass, and would favor a more 
efficient $Ly\alpha$ photon destruction operated by dust. The presence
of dust and gas would also suggest that the escape fraction of 
ionizing radiation is very low, as it has been observed in $L^{\star}$ 
star-forming galaxies at $z\simeq 3$ \citep{vanz10a, boutsia11}.

Second, {\it GDS\_1408}  is one of the
more extended sources among the $z\sim 7$ candidates (half light
radius of $0.26''$ , i.e., 1.4 proper Kpc, \citealt{grazian11}, with an
elongated morphology of $4.8\times 2.5$ proper Kpc, see Figure~\ref{HST}).
It has been shown that the $Ly\alpha$ equivalent width and the
size observed at the 1500\AA~rest-frame of the stellar
continuum anti-correlate, such that on average the emitters appear 
more compact and nucleated than the non-emitters, 
with average half light radius $\lesssim 1$kpc
\citep{law07,vanz09,pente10}.

{\it GDS\_1408} is moderately star-forming and contains a non
negligible amount of dust, it is UV bright and spatially extended.
While we cannot exclude that the CGM/IGM plays an additional role,
  the properties of {\it GDS\_1408} are consistent with those of others
  lower-z $L^{\star}$ star-forming galaxies that show faint
  $Ly\alpha$ emission, and is not a ``smoking gun'' of an increased
  neutral IGM at $z=7$. It is worth noting that we are discussing
  a single galaxy case which clearly cannot be considered as representative
  of a population. The deficit of $Ly\alpha$ lines from redshift 6 to 7 
  recently observed is significant and is based on a statistical analysis that 
  compares tens of star-forming galaxies selected with very similar color techniques,
  indepenently from the presence of the $Ly\alpha$ line.
  However, as noted also in \citet{schenker14}, galaxies like 
  {\it GDS\_1408} are not the best tracers of an IGM damping.
  The higher is the probability the $Ly\alpha$ emission is internally absorbed,
  the lower is the power of tracing the neutral gas fraction of the 
  circum/inter-galactic medium.
 
  Nonetheless, it is important to assess the nature of 
  ``continuum-only'' star-forming galaxies at $z>6.5$, still an unexplored
  line of research.
  As reported in this work, the tentative spectroscopic investigation of 
  ``continuum-only'' galaxies at $z>6.5$ shows all the limitations of the
  8-10 meter class telescopes coupled with optical spectroscopy ($\lambda<1.1\mu m$).
  Different and future facilities are needed to shed light on their nature. 

\subsection{Future prospects: ALMA, JWST and ELT}

We have shown the current limits of 8-10m class telescopes in 
the spectroscopic characterization and redshift measurement of non-$Ly\alpha$
emitters at $6.6<z<7.3$. If $z\sim 7$ is a critical value above which the 
visibility of $Ly\alpha$ lines decreases drastically, then future facilities are
necessary to capture the UV continuum-break and the ultraviolet absorption
lines and/or optical nebular emission lines at $7<z<10$.
JWST-NIRSPEC will probe the typical nebular emission
lines, e.g., [OII]3727, H$\beta$, [OIII]4959-5007 up to
$z\simeq 9$ ($5 \mu m$) and the extremely large
telescopes (ELT, 30-40m diameter) will allow us to cover the mid-infrared 
part (e.g., EELT-METIS, 8-14 $\mu m$) and to probe the UV continuum with
S/N=10 down to $J\simeq 27$ and therefore to study in detail the ultraviolet
absorption lines, not to mention the possibility to perform high spatial resolution analysis.

\begin{figure}
\centering
\includegraphics[width=8.5 cm, angle=0]{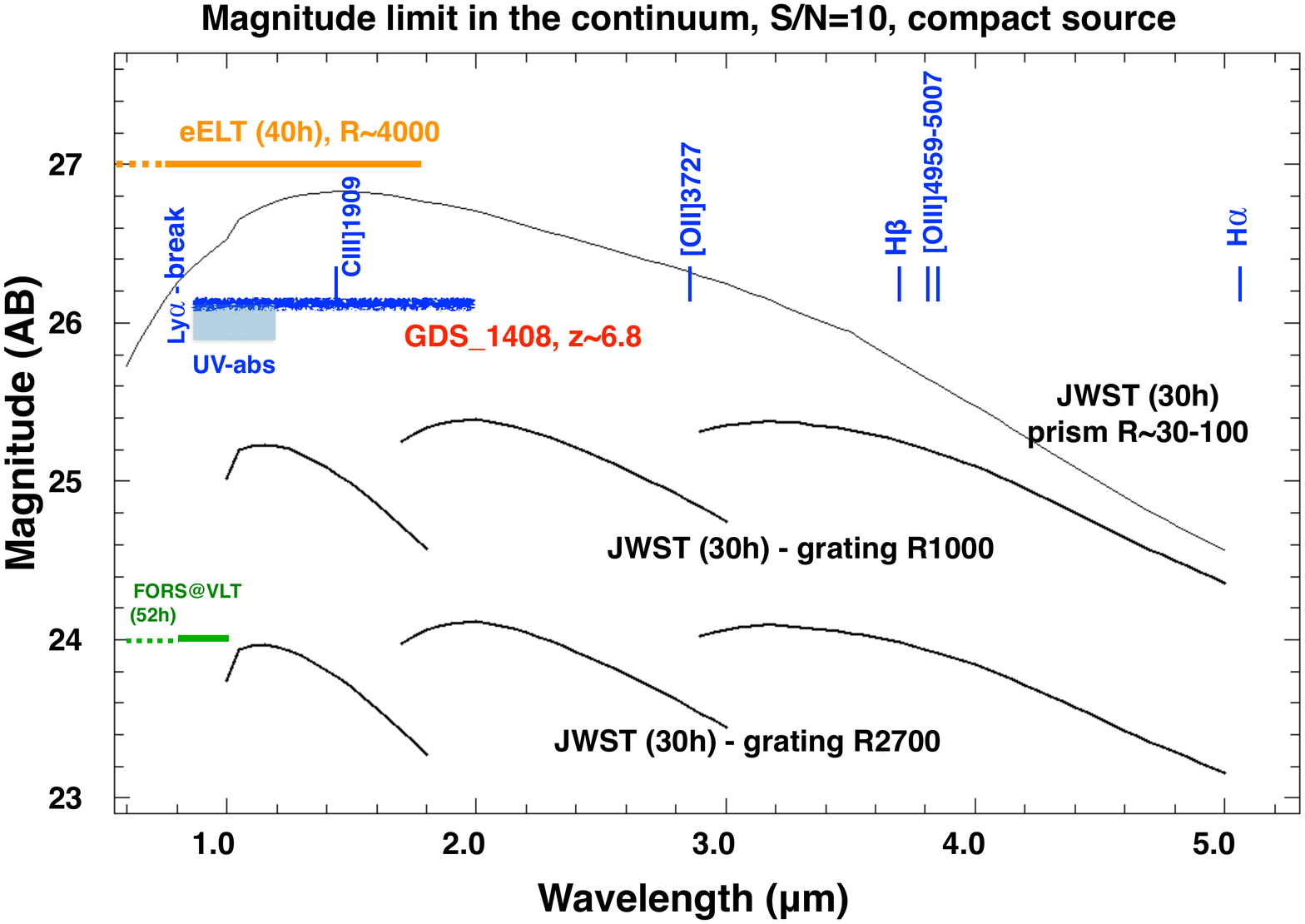}
\caption{Expected magnitude limits in the continuum (S/N=10) with
  30-50h integration time are shown for ELT, (orange, Evans et
  al. 2013), JWST (black, available in tabular form 
{\it http://www.stsci.edu/jwst/instruments/nirspec/sensitivity/})
 and VLT-FORS (green, this work). The estimated limit for KMOS is
magnitude $\simeq 21.5$ in the YJ band with 50h integration 
(rescaling from the KMOS manual, Sect. 2.3.4).
The schematic position of {\it GDS\_1408} is reported (blue), with the
continuum magnitude $\simeq26.1$, the typical ultraviolet absorption 
lines from the $Ly\alpha$ to CIV~1550 (grey region, UV-abs) and the basic 
emission lines, e.g., CIII]1909, [OII]3727, H$\beta$,
[OIII]4959-5007 and H$\alpha$.
}
\label{FUTURE}
\end{figure}

In particular, the case of {\it GDS\_1408}  is shown in
Figure~\ref{FUTURE}, where a schematic view of the limits on
the continuum at S/N=10 are reported for the ELT, JWST and VLT telescopes.
A good characterization of the ultraviolet absorption lines will be
feasible with the ELT.  JWST would marginally identify the trace of
the continuum ($S/N<5$), but will open for the measure of optical
emission lines (up to [OIII]4959-5007).

While JWST and ELT will definitely perform these kind of studies,
at present the measure of the redshift with 8-10m telescopes could
be achieved looking at emission lines different from $Ly\alpha$.
Though unusual, lines like OIII]$\lambda\lambda$1661-1666 and 
CIII]$\lambda\lambda$1907-1909 could be identified by means of
near infrared spectrographs. Moreover, depending on the source of 
ionizing photons, other lines like NV~1240, NIV]$\lambda\lambda$1483-1486, 
CIV~1550, HeII~1640 can also be measured (e.g., \citealt{ost06,vanz10b, raiter10, steidel14}).

Another promising facility that might be able to determine the
spectroscopic redshift of this galaxy is ALMA.
Indeed, a spectral scan of the [CII]158$\mu m$ line encompassing
  the full 2-sigma range of uncertainty could be covered with two ALMA
  spectral bands. Assuming that the local SFR-[CII]158$\mu m$ relation of
  \citet{sarg12} holds also at these high redshifts, based on the
  best-fitting SFR=20$M_{\odot}/$yr$^{-1}$ we expect a flux of
  0.24 Jy km/s, that can be secured in a relatively short time 
  ($<6$ hours in the Cycle2 sensitivity) with ALMA.
A molecular line scan that covers the 3mm window (89-115 GHz) can also reveal multiple CO 
transitions in {\it GDS\_1408}.  
In particular at $z>6.5$ the transitions $J=7\rightarrow6$ and $J=6\rightarrow5$ 
(where J is the rotational quantum number) are observable.
Given the UV slope and the star formation activity, and assuming the
relation between CO line luminosity and SFR \citep[see][]{carilli13, decarli14},
the expected CO transitions can be detected ($S/N \gtrsim 3$) with an ALMA 
3mm scan with sensitivity limit of $0.1$mJy~beam$^{-1}$.

\section{Conclusions}\label{conc}
We have reported the combined VLT/FORS2 spectroscopy of one of the most 
reliable $z\simeq 7$ galaxy candidates in the Hubble Ultra Deep Field. 
Three different programs at VLT targeted {\it GDS\_1408} in the last five years, 
for a total integration time of {\it 52h}. We have retrieved from the 
ESO archive all these data, and re-analyzed them in a consistent way. 
Our main results are:

\begin{enumerate}

\item{{\bf An upper limit on the Ly$\alpha$ emission:} We are able to
  place a stringent upper limit of $f$($Ly\alpha$)~$<3 \times
  10^{-18}~erg/s/cm^{2}$ at 3-sigma (in the sky-lines) and up to 9-sigma
  (in sky-free regions) in the whole wavelength range, i.e., up to
  $z=7.0$ (see Figure~\ref{SIM_CONT}). This corresponds to an upper
  limit on the equivalent width $EW<9$~\AA, with the same statistical
  accuracy. With the deeper data used here we do not confirm the
  tentative detection of a weak line located at $\lambda = 9691.5 \pm
  0.5$~\AA, corresponding to a redshift of 6.972, that we claimed in
  F10.}

\item{{\bf Redshift refinement:} The combination of ultradeep spectroscopy,
superb HST information and narrow band imaging, have allowed us to refine 
the photometric redshift value of {\it GDS\_1408}, that turns out to be $6.82\pm0.1$. 
The same analysis indicate that {\it GDS\_1408} has a nearly solar
metallicity and is relatively dust attenuated ($A_{1600}\simeq 1$) galaxy. }

\item{{\bf $Ly\alpha$ escape fraction:} 
We derived a $f_{esc}^{eff}(Ly\alpha)<8$\%.
The $Ly\alpha$ attenuation can be a combination of internal (ISM) and external
effects (IGM).}

\end{enumerate}

Even though we cannot rule out a major contribution of the IGM in 
damping the line, the {\it most} plausible interpretation is that {\it G2\_1408} is a 
star-forming galaxy at $z\simeq 6.82$ moderately evolved and containing sufficient 
gas and dust to attenuate the $Ly\alpha$ emission, before it reaches
the intergalactic medium. 
If compared with fainter ($0.05L^{\star}<L<0.2L^{\star}$), young, less massive 
and less evolved counterparts that show $Ly\alpha$ emission and much steeper 
UV slopes $\beta < -2.5$ (e.g., \citealt{balestra13,vanz14}),
{\it G2\_1408} appears more massive and evolved ($\beta \simeq -2$).

An absence of the line due to dust absorption is not 
in contrast with recent results on the deficit of $Ly\alpha$ lines ascribed 
to a possible increase of the neutral gas fraction of the
IGM between $z\sim6$ and 7 (e.g., P11, P14, \citealt{fink13}).
In fact, the apparent drop of lines is an evidence based on statistical 
samples, in which the properties of redshift 6 and 7 galaxies are compared 
in a differential way, and applying similar color selection techniques.

Redshift 7 sources like {\it GDS\_1408}, and in general even fainter ones at $J=27$, 
will be well studied spectroscopically with ELT and JWST telescopes, in particular 
for what concerns the UV absorption lines and nebular (optical) emission lines,
respectively. 
As an example, JWST will provide the systemic redshift from rest-frame optical nebular 
emission lines (Oxigen, Balmer lines) and the ELT will capture the signature of the
ISM in absorption in the rest-frame ultraviolet.
This will be feasible not before 2018-2020.
Currently, a facility that can provide redshift measures at $z\gtrsim 7$ and characterize 
the properties of the ISM and stellar population is ALMA. 
Sources like {\it GDS\_1408} with a reliable guess on the
redshift value are suitable candidates for the scanning mode of ALMA.

\begin{acknowledgements}
    We thanks R. Bouwens for providing information about the deep IRAC magnitudes of the
    source and P. Oesch for providing information about the {\it GDS\_1408} target
    in the spectroscopic program 088.A-1008(A). We thank F. Vito, M. Mignoli, A. Cimatti and
    S. de Barros for useful discussions. Part of this work has been funded through
    the INAF grants (PRIN INAF 2012).
\end{acknowledgements}


\end{document}